\documentclass{emulateapj}

\usepackage{amssymb,amsmath}
\usepackage[dvips]{epsfig,color}

\newcommand{\eg}{e.g.}

\newcommand{\etal}{et al.}

\shorttitle{Triaxiality Inhibitors}
\shortauthors{Barnes \& Dowling}

\begin{document}

\title{Triaxiality Inhibitors in \textit{N}-Body Simulations}
\author{Eric I. Barnes}
\affil{Department of Physics, University of Wisconsin---La Crosse, La
Crosse, WI 54601}
\email{barnes.eric@uwlax.edu}
\author{Evan Dowling \altaffilmark{1} }
\affil{Department of Physics, University of Wisconsin---La Crosse, La
Crosse, WI 54601}
\altaffiltext{1}{University of Maryland, College Park, MD 20742}
\email{edowling@terpmail.umd.edu}

\begin{abstract}

Numerous previous studies have investigated the phenomenon wherein
initially spherical \textit{N}-body systems are distorted to triaxial
shapes.  We report on an investigation of a previously described
orbital instability that should oppose triaxiality.  After verifying
the instability with numerical orbit integrations that extend the
original analysis, we search for evidence of the instability in
\textit{N}-body systems that become triaxial.  Our results highlight
the difficulty in separating dynamical process from finite-\textit{N}
effects.  While we argue that our analysis points to the presence of
the instability in simulated triaxial systems, discreteness appears to
play a role in mimicking the instability.  This suggests that
predicting the shapes of real-world systems, such as dark matter halos
around galaxies, based on such simulations involves more uncertainty
than previously thought.

\end{abstract}

                          
\section{Introduction}\label{intro}

Elliptical galaxies, galactic bulges, and some star clusters are
well-approximated by collisionless models.  It is also commonly
assumed that dark matter halos around galaxies are best described as
collisionless systems.  All of these systems commonly have triaxial
shapes that are typically described in terms of semi-axis
lengths labeled $a$, $b$, and $c$ for long, intermediate, and short
axes, respectively.  In particular, simulated dark matter halos that
form via hierarchical growth in cold dark matter cosmologies are
generally triaxial with mean axial ratios $b/a \sim 0.6 $ and $c/a
\sim 0.4$ \citep{betal86,be87,fetal88,dc91,js02,bs05,aetal06}.

Attempts to understand the development of this triaxiality typically
invoke what has become known as the radial orbit instability (ROI)
\citep[\eg,][]{ma85,betal86,pp87}.  This instability starts with a
spherical system that hosts particles on radial orbits.  The radial
anisotropy of the orbits may be prescribed by initial conditions or it
may develop as a result of a dynamically cold collapse.  Detailed
comparisons of the onset and triggering of the ROI in these situations
has been previously presented \citep*{blw09}.  Independent of the
specifics of the initial conditions, the instability grows as
neighboring orbits tend to become aligned along a direction that has
slightly more particles than others.  The end products of this
instability are important as the shapes of simulated dark matter halos
impact how they will interact with satellite objects
\citep[\eg,][]{setal10}.  At the same time, this instability also
affects density profiles of simulated halos \citep{betal05,betal08}.
Both of these effects could have observable consequences, if real
halos are accurately modeled by simulations.

Our goal in this work is to complement those earlier studies by
investigating how simulated systems halt, and even reverse, their
evolutions to triaxial shapes.  In particular, we are interested in an
orbit instability that arises in triaxial systems which deflects
orbits that lie near principal planes.  This type of instability has
been considered previously \citep{b81,mf96}, but \citet{aetal07} have
provided a more recent detailed analysis.  What we refer to as the
Adams instability acts to change orbits that support triaxial shapes.
Particles that orbit near principal planes can show exponential
divergence perpendicular to those planes.  In this way, orbital
families that would underlie a triaxial shape are depleted.  We expect
that a system vulnerable to the Adams instability would become more
spherical or, at least, have a limited range of non-sphericity.  The
work in \citet{aetal07} focuses on the behavior of orbits in the
central region of a cusped potential.  As we are interested in looking
at the behavior of entire simulated systems, our work picks up a
thread from theirs by extending the investigation to outer regions of
triaxial, cuspy potentials.

The larger part of our investigation searches for evidence of the
instability in \textit{N}-body simulations.  We use the publicly
available {\small GADGET-2} code \citep{s05} to evolve systems with
$10^5 \le N \le 10^6$ in cosmological and non-cosmological situations.
The time dependence of the instability for orbits in smooth,
analytical potentials forms a template to search for signs of the
instability in \textit{N}-body systems, however the discrete nature of
these systems complicates our approach.  A similar situation arises in
recent investigations of isolated and initially cold systems which
show that initially spherical systems with small-scale density
fluctuations will evolve in a manner similar to systems with initial
triaxiality \citep{bs15}.

The remainder of the paper will lay out our methods and results.
Section~\ref{adams} details our extension of the \citet{aetal07} work
to include non-central regions of triaxial systems.  Descriptions of
the initial conditions and evolutions of our \textit{N}-body models
form \S~\ref{sims}.  Our investigation of the impact of discreteness
effects follows in Section~\ref{discrete}.  We summarize the results
of our search for evidence of the Adams instability in \textit{N}-body
simulations in \S~\ref{conclude}.

\section{Adams Instability in Smooth Potentials}\label{adams}

The Adams instability has been observed in triaxial potentials with
the form
\begin{equation}\label{triphi}
\Phi(x,y,z)=2\int_0^{\infty} \frac{\psi(m) \,
du}{\sqrt{(a^2+u)(b^2+u)(c^2+u)}},
\end{equation} 
where $m$ is length defined by
\begin{equation}
m^2=\frac{x^2}{a^2} + \frac{y^2}{b^2} + \frac{z^2}{c^2}.
\end{equation}
The values $a$, $b$, and $c$ specify the long, intermediate, and short
semi-axis lengths of the triaxial shape.
The $\psi$ function is defined as \citep{c69,bt87}
\begin{equation}
\psi(m)=\int_m^{\infty}2m^{\prime} \, \rho(m^{\prime}) \, dm^{\prime}. 
\end{equation} 
The \citet{aetal07} work uses two common analytical density profiles,
\citep*{h90,nfw96}
\begin{eqnarray}
\rho_{NFW} & = & \frac{\rho_0}{m(1+m)^2} \quad \mbox{and} \nonumber \\
\rho_{H} & = & \frac{\rho_0}{m(1+m)^3}.
\end{eqnarray} 
As the precise form of the potential is not of interest here, we will
concern ourselves only with the Navarro-Frenk-White (NFW) form and set
$\rho_0=1$.

\begin{figure}
\scalebox{0.5}{
\includegraphics{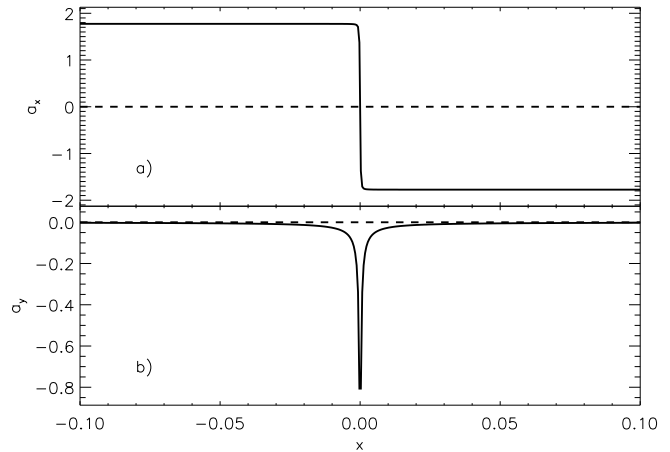}}
\caption{The $x$ dependence of accelerations for a potential with
$a=\sqrt{2}$, $b=1$, $c=0.5071$ along a path where $y=z=10^{-4}$.
Panel a shows the $x$-component of the acceleration while panel b
shows the $y$-component.  The rapid change in sign of
$x$-acceleration coupled with the more impulsive nature of the
$y$-acceleration lead to strong deflections of particles moving along
principal axes as they pass near the origin.
\label{afig}}
\end{figure}

\begin{figure}
\scalebox{0.5}{
\includegraphics{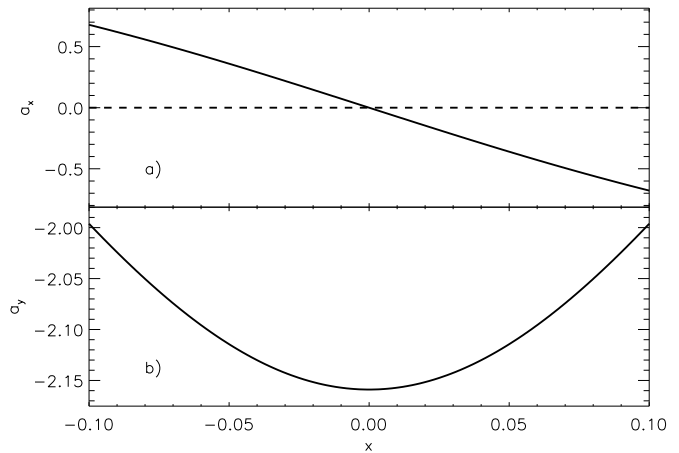}}
\caption{The $x$ dependence of accelerations for a potential with
$a=\sqrt{2}$, $b=1$, $c=0.5071$ along a path where $y=0.2$ and
$z=10^{-4}$.  Panel a shows the $x$-component of the acceleration
while panel b shows the $y$-component.  The step-like feature of the
$x$-acceleration and the impulsive nature of the $y$-acceleration have
been removed, guaranteeing that particle velocities do no undergo
sudden changes.
\label{afig2}}
\end{figure}

Analytical expressions for the potential and acceleration in the
central regions of such density distributions have been derived
\citep{pm01,aetal07}.  Based on these expressions, one can focus on
the behavior of accelerations for particles that approach the origin
along one of the principal axes.  For concreteness, we imagine a
particle moving towards the center nearly along the long ($x$) axis
with the particle's $y$ and $z$ positions being $\ll 1$.  With these
conditions, the $x$-acceleration becomes a step function about the
origin (Figure~\ref{afig}a).  The acceleration is constant and
negative for $x>0$ and immediately becomes positive (and constant) for
$x<0$.  The $y$- and $z$-accelerations do not change sign with $x$,
but reach their largest magnitudes at $x=0$.  A representative
$y$-acceleration curve is shown in Figure~\ref{afig}b.  These
accelerations work to create a large change in the direction of the
velocity vector perpendicular to the long axis as it passes near the
origin.  This seems to be the origin of the instability.  It is absent
for particles that do not move along principal axes because the
accelerations take on very different characteristics in other regions.
Figure~\ref{afig2} shows acceleration behaviors near $x=0$ when $z$
remains small, but $y$ does not.  In this non-principal axis
situation, the $x$-acceleration is nearly harmonic and the
$y$-acceleration is nearly constant.

\begin{figure}
\scalebox{0.5}{
\includegraphics{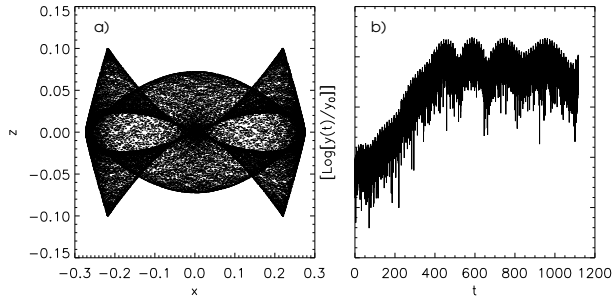}}
\caption{Panel a shows a projection of a particle orbit that
originates near the $x-z$ plane, with initial $y=10^{-8}$ and
undergoes exponential growth in a potential with $a=\sqrt{2}$, $b=1$,
$c=0.5071$.  This shape should be compared to the orbit in Figure 3 of
\citet{aetal07}.  Panel b shows how the $y$ position of the orbit
grows with time, replicating the evidence of the instability shown in
Figure 4 of \citet{aetal07}.  The exponential growth of the deviation
from the initial orbit plane results in the linear growth segment.
The time units are arbitrary, but typical orbits investigated have
periods $1 \lesssim T \lesssim 10$. 
\label{valid}}
\end{figure}

\begin{figure}
\scalebox{0.4}{
\includegraphics{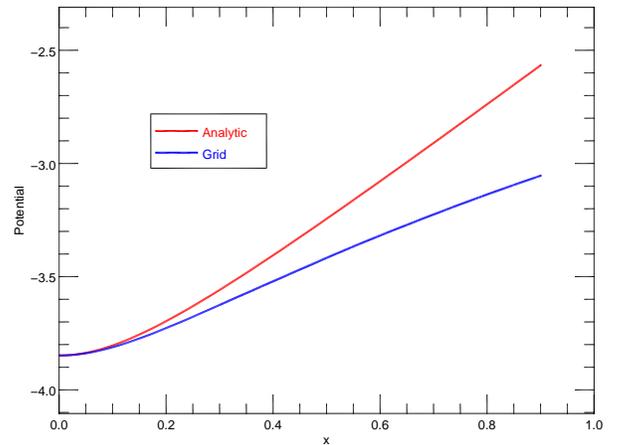}}
\caption{Representative potential values along the major axis of a
triaxial model.  The red line indicates how the centrally-limited
approximation would behave beyond its region of validity.  The blue
line represents the full NFW potential.  We investigate orbits that
move through regions where the difference between the potentials is
significant.
\label{potcomp}}
\end{figure}

We begin by validating our numerical orbit integration routines
through comparisons with orbits highlighted in \citet{aetal07}.  The
integration scheme uses a standard variable time step, Runge-Kutta
approach \citep{petal94}.  Using the analytical approximation to the
potential near the center, we have found a good match to an orbit in
\citet[][their Figure 3]{aetal07}.  The close comparison of orbit
shape and size, along with the instability behavior, between
Figure~\ref{valid} and similar figures in \citet{aetal07} serve as
evidence that we are seeing the same behavior using our techniques.
With the success of this base step, we extend to a full NFW potential,
not relying on the limiting, analytical forms.  Figure~\ref{potcomp}
shows a comparison between long-axis potential values from the
centrally-limited approximation (labeled `Analytic') and the full NFW
form (labeled `Grid').  In extending beyond the central region,
Equation~\ref{triphi} has been numerically evaluated to determine
potential values on a three-dimensional grid with logarithmic spacing.
Similarly, expressions for accelerations have been analytically
determined by differentiating Equation~\ref{triphi}.  The resulting
integrals have then been numerically evaluated on the same grid as the
potentials.  The logarithmic grid allows for better spatial resolution
near the center of the potential where more dramatic changes in
potential and acceleration occur.  Spline interpolation routines are
used to determine values at arbitrary locations in the grid.  We
utilize the {\small EZSPLINE} implementation of the {\small PSPLINE}
library, {\tt http://w3.pppl.gov/ntcc/PSPLINE/}.  In regions where
orbits are investigated, analytical and grid-based values of potential
vary by as much as 15\%.  Accelerations are calculated on the same
grid by numerically integrating spatial derivatives of
Equation~\ref{triphi}.  With these values, our integrations conserve
energy to roughly one part in one thousand over roughly 100 orbits.

\begin{figure}
\scalebox{0.5}{
\includegraphics{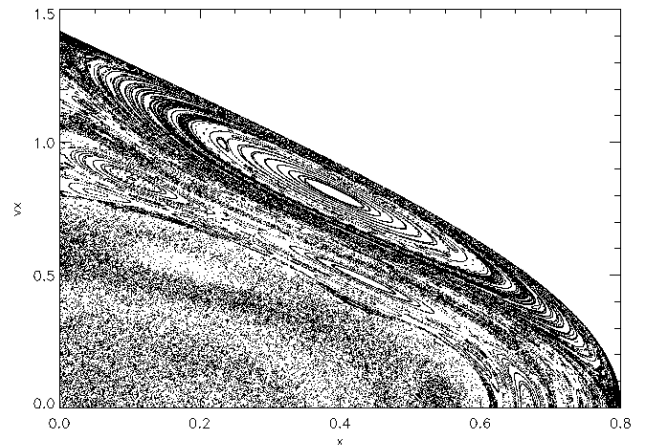}}
\caption{Analogous to the surface of section plot in Figure 1b of
\citet{aetal07}, this plot shows a slice through phase space that
clearly delineates regions of regularity (smooth, connected curves)
from irregularity.  The axis ratios used match those in
\citet{aetal07} (3:4:5), but the full NFW potential is utilized.  Each
orbit has $|y|_{\rm initial} \le 10^{-8}$ but varying $x$ and $z$
initial positions.
\label{gridsos}}
\end{figure}

With the full potential, we have generated surface of section plots to
study a larger family of possible orbits.  Figure ~\ref{gridsos} shows
a surface of section plot for orbits with initial $y$ values $\approx
10^{-8}$ but varying initial $x$ and $z$ locations. Stable orbits form
regular curves whereas unstable orbits appear as swarms of scattered
points throughout regions of a surface of section plot.  Orbits in the
inner regions most typically show evidence of the instability, but
there are also unstable regions for orbits which pass through areas
where the difference between the approximate and full potentials seen
in Figure~\ref{potcomp} becomes significant.  As a result of this
extension to previous work, we conclude that the Adams instability
should be present in any triaxial system with a cuspy inner profile,
independent of the details of the potential further from the center.
As cuspy cores are common results in \textit{N}-body simulations of
collapsing systems, we now turn to finding evidence of the Adams
instability among them.

\section{\textit{N}-body Simulations}\label{sims}

One of the main results of the \citet{aetal07} work is the recognition
that there is a specific signature of the instability.  As an orbit
experiences the instability, its position perpendicular to the
principal plane in which it originally moved exponentially increases.
We search for evidence of this behavior among the orbits that compose
various \textit{N}-body systems.

We have evolved systems with both $N=10^5$ and $N=10^6$ particles.
All systems are initialized by randomly placing particles in spherical
volume according to a prescribed density distribution.  Typically, we
will discuss models with Gaussian ($\rho \propto e^{-r^2}$) density
distributions, but models with cuspy ($\rho \propto 1/r$) profiles
have also been evolved.  Past work has indicated little difference
between the onset of the ROI in such systems when the amount of
initial kinetic energy present is limited to induce triaxiality
\citep{blw09}.  For non-cosmological evolutions, this amounts to
limiting the initial virial ratio of the system.  We define the virial
ratio as $Q=2T/|W|$, where $T$ and $W$ are the kinetic and
gravitational potential energies, respectively.  Velocity
distributions are made initially isotropic by randomly orienting each
particle's velocity.  Cosmological evolutions are started
somewhat differently.  The random component of the kinetic energy is
limited and each particle's initial velocity is a combination of
Hubble expansion and isotropic random motion.

All \textit{N}-body evolutions have been calculated using the publicly
available {\small GADGET-2} code \citep{s05}.  Simulations with
$N=10^5$ use softening values $\delta = 10^{-4}$.  In $N=10^6$
simulations, evolution times are kept reasonable by adopting softening
length values $\delta \propto N^{-1/2}$ \citep{petal03}.
Specifically, we set $\delta = 4 \times 10^{-3}$.  Test runs with
different values of $\delta$ have been performed and produce
essentially identical results.  Cosmological evolutions run between
redshifts of nine and zero, while non-cosmological evolutions proceed
for at least ten initial-system crossing times.  The virial ratio
behaves reasonably, with early oscillations settling down to the
equilibrium value of one by the end of an evolution.

\begin{figure}
\scalebox{0.5}{
\includegraphics{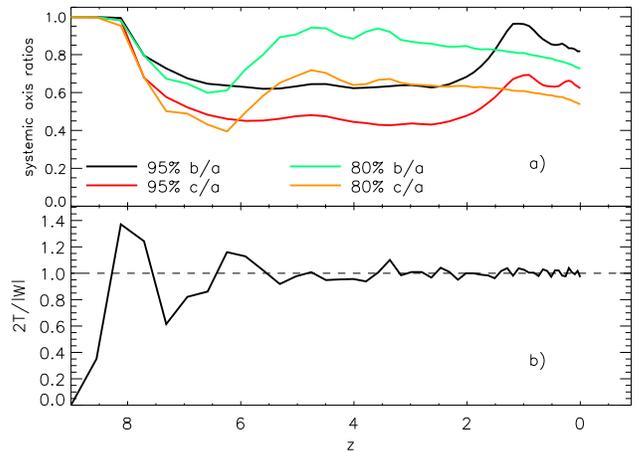}}
\caption{Results from a cosmological evolution of a system with
$N=10^6$, Gaussian initial density distribution, and zero initial
random kinetic energy.  Panel a shows the behavior of systemic
axis ratios for two mass cutoff values.  Using the innermost 80\% of
the mass in a system to determine axis ratios excludes many of the
particles that are either escaping or near-escape as a result of such
a cold collapse.  Panel b illustrates the evolution of the virial
ratio.  This rapid approach to equilibrium is essentially the same for
all of our evolutions.
\label{arevst}}
\end{figure}

Periodic output files are analyzed in the post-evolution stage.
Particle positions and velocities are used to determine axis ratio and
energy behaviors as functions of time/redshift.  The $x$, $y$, and $z$
axes are defined as the long, intermediate, and short axes of the
system determined at the last output time.  In cosmological
simulations, all coordinates and velocities are comoving.  All
previous output values are transformed to this coordinate system for
consistency.  Examples of analysis products are presented in
Figure~\ref{arevst}.  In general, our initially dynamically cold
systems quickly deform to triaxial shapes followed by a longer-term
relaxation towards sphericity.

\begin{figure}
\scalebox{0.5}{
\includegraphics{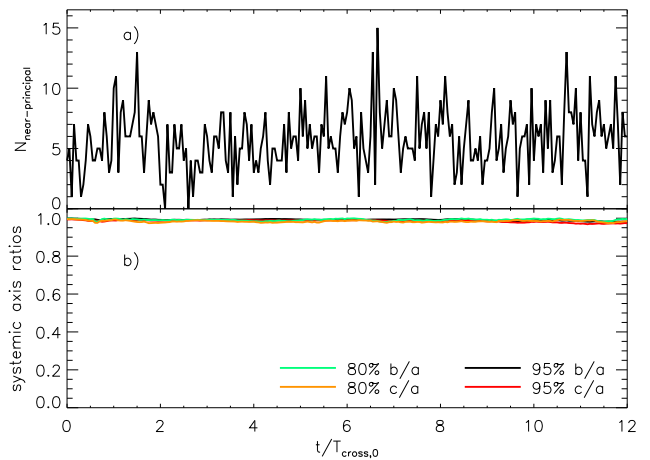}}
\caption{Results from a non-cosmological evolution of a system with
$N=10^5$, Gaussian initial density distribution, and just enough
kinetic energy to avoid becoming triaxial (initial $Q=0.08$).  Panel
a shows the behavior of the number of particles in near-principal
plane orbits.  The near constancy of this value is seen in all of our
non-cosmological evolutions, even those that result in triaxiality.
Panel b illustrates the systemic axis ratios in the same way as in
Figure~\ref{arevst}.  This model remains essentially spherical
throughout its evolution.
\label{ansphere}}
\end{figure}

We also filter for particles that are on orbits that should be
susceptible to the Adams instability.  Focusing on particles that are
near the $x-z$ plane with small $y$-velocities gives us a subset that
may show exponential growth perpendicular to the long axis of the
system.  Through trial-and-error, we have determined ranges of
$y$-positions and velocities that balance the need to only look at
particles with near-principal plane orbits with the desire to have as
many particles as possible to analyze.  Specifically, we demand that
particles have $|y|\le 10^{-4}$ and $|v_y|\le 5\times 10^{-2}$ to be
investigated further.  We note that in varying these values, there are
no changes to the qualitative behaviors we will now discuss.  Simply
tracking the numbers of particles that fit these constraints should
provide a crude way to distinguish between the presence and absence of
the instability.  In the absence of the instability we expect the
numbers of near-principal plane particles to be relatively steady.  An
example of this kind of behavior is shown in Figure~\ref{ansphere}.
This figure corresponds to a non-cosmological simulation where there
is enough initial kinetic energy to prevent triaxiality.

\begin{figure}
\scalebox{0.5}{
\includegraphics{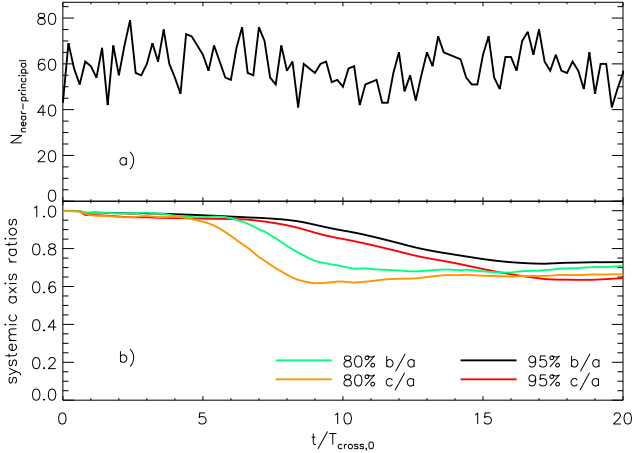}}
\caption{Results from a non-cosmological evolution of a system with
$N=10^6$, Gaussian initial density distribution, and not enough
kinetic energy to avoid becoming triaxial (initial $Q=0.04$).  The
panels contain the same information as those in Figure~\ref{ansphere}.
We note the same constant behavior of the number of
near-principal-plane orbits even in the face of significant
triaxiality in the system.
\label{antnc}}
\end{figure}

\begin{figure}
\scalebox{0.5}{
\includegraphics{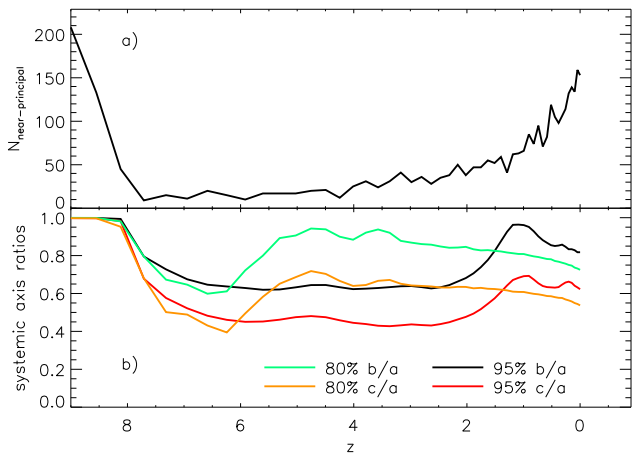}}
\caption{Results from the same evolution shown in Figure~\ref{arevst}.
Panel a shows the behavior of the number of particles in
near-principal plane orbits for a system.  Unlike the analogous panel
in Figure~\ref{ansphere}, there are two different trends in this
behavior.  There is a sharp initial decrease in the particles trapped
near the principal plane.  This decrease is seen only in systems with
zero initial random kinetic energy (independent of cosmological
expansion).  Also, as redshift of zero is approached, this number
climbs significantly.  This late-time climb is common among all of the
cosmological simulations we have run, independent of the duration
and/or severity of the triaxiality induced.  Panel b illustrates the
corresponding systemic axis ratios.  With the initial conditions we
have investigated, all cosmological evolutions result in triaxiality.
However, increasing the amount of initial random kinetic energy delays
the onset of triaxiality.
\label{antriax}}
\end{figure}

Our hypothesis is that an active instability should depopulate
near-principal plane orbits.  As an active instability requires
triaxiality, we present results from a simulation where the initial
conditions are cold enough that the ROI will act.  Figure~\ref{antnc}
is analogous to Figure~\ref{ansphere}, but for a colder,
non-cosmological system.  Counter to our expectations, there is no
obvious evolution of the number of near-principal-plane orbits.
Another example of an evolution of near-principal-plane particles is
shown in Figure~\ref{antriax}.  In this cosmological simulation, there
is no initial random kinetic energy.  The loss of near-principal plane
orbits begins even as the system is still mostly spherical.  When the
system is most triaxial, the population seems rather stable.  As the
system moves towards a more spherical shape, the population rises
again.

In other cosmological simulations with non-zero initial random kinetic
energy, plots analogous to Figure~\ref{antriax} lack the large initial
value and subsequent decrease.  The absence of any initial random
velocities produces a sizeable population of orbits that satisfy our
criteria for being near-principal plane, but subsequent evolution
leads to random motion that remove them from consideration.  Similar
large decreases are also present in non-cosmological evolutions with
zero initial kinetic energy.  Apparently, the decrease seen in
Figure~\ref{antriax} has nothing to do with the Adams instability.  In
the dynamically warmer cosmological simulations, triaxiality takes
longer to develop, but the increase in near-principal-plane particles
as $z=0$ is approached remains.  Similar late-evolution increases are
not present in any of our non-cosmological simulations.  The cause of
this difference is not clear at present.

\begin{figure}
\scalebox{0.5}{
\includegraphics{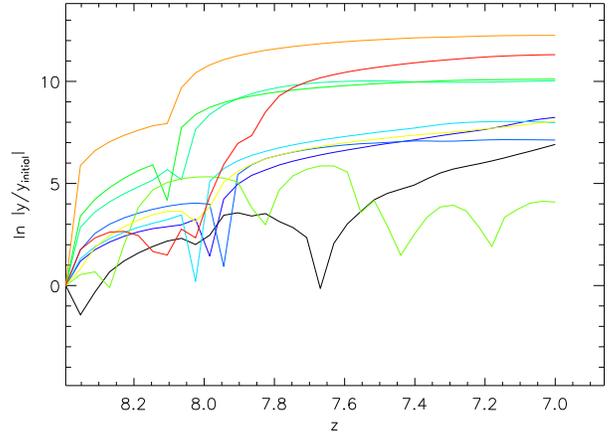}}
\caption{Behavior of the $y$-positions of several particles in the
same \textit{N}-body simulation as in Figure~\ref{antriax}.  The
starting point has been chosen so that it roughly agrees with the time
when the system is beginning to become triaxial.  Some orbits share the
same qualitative behavior as Adams unstable orbits.
\label{nblogy}}
\end{figure}

We have also searched for a signal of the Adams instability in the
behavior of individual orbits of masses in the \textit{N}-body
simulations.  After identifying particles in near-principal plane
orbits at a specified point in an evolution, we have tracked the
subsequent positions of those particles for the remainder of the
simulation.  Plots of near-principal plane orbit behaviors are shown
in Figure~\ref{nblogy}.  At least some of the orbits show the same
qualitative behavior as unstable orbits in smooth potentials.
Numerous particles experience rapid rises in their positions
perpendicular to the principal plane followed by a steady-state
behavior. Unfortunately, a similar plot can be created starting at a
point in time when the system is essentially spherical.  The large
changes in $y$-positions appear to happen even though the potential
should not support the Adams instability.  Without an unambiguous
signal like the one shown in Figure~\ref{valid}b, and given that
instability-like behavior occurs in spherical systems, the analysis of
these \textit{N}-body systems does not provide conclusive evidence of
the Adams instability.

\section{Investigating Discreteness Effects}\label{discrete}

The qualitative hints of the instability discussed above do
provide some motivation to look beyond our simplest approaches.  Given
that there can be numerical effect competing with physical processes,
we next turn to untangling any instability effects from
finite-\textit{N} effects by taking parallel tracks.  We investigate
how small-scale noise superimposed on a smooth background potential
impacts the Adams instability.  Separately, we analyze orbits in
smoothed potentials derived from \textit{N}-body simulations.  The
results of these investigations are described below.

Before moving on, we note that this discussion is not intended to
imply that these simulations suffer from global-scale two-body
effects.  In previous work with similar models, we have found that
neither mass segregation nor pervasive impulsive events occur
\citep{blw09}.  By these yardsticks, our simulations are
collisionless.

We have added Gaussian perturbations to the smooth potentials
described in Section~\ref{adams}.  Specifically, each perturbation has
the form,
\begin{equation}
\Phi_{\mathrm{pert}}(\mathbf{r})=A\exp{\left[ 
\frac{-(\mathbf{r}-\mathbf{r}_0)^2} {2\sigma_p^2} \right]},
\end{equation}
with $A$ controlling the strength of the potential and $\mathbf{r}_0$
specifying its location.  The width of the Gaussian $\sigma_p$ has been
set to half the average distance between perturbers.  Typically,
thousands of perturbers are placed randomly throughout a potential.
Using the orbit integration scheme described in Section~\ref{adams},
we have investigated the same range of initial conditions for these
bumpy potentials.

\begin{figure}
\scalebox{0.5}{
\includegraphics{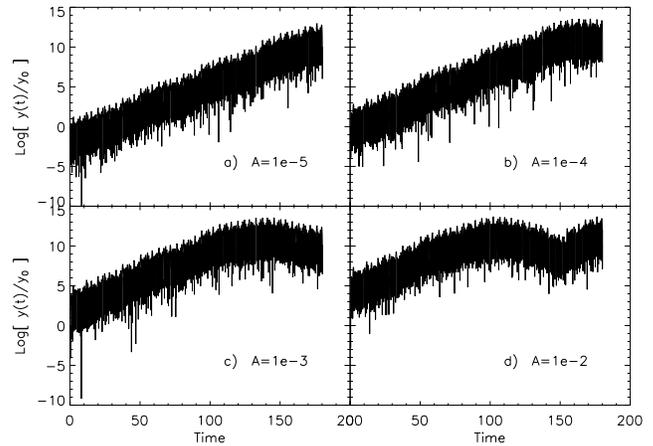}}
\caption{Behaviors of an unstable orbit in a triaxial potential with
$a=5$, $b=4$, $c=3$ and $|y|_{\rm initial} <10^{-8}$.  Panels a
through d correspond to perturbation strengths $10^{-5}$, $10^{-4}$,
$10^{-3}$, and $10^{-2}$, respectively.  The tell-tale exponential
growth in the unstable $y$ direction is independent of the perturbing
strength.  However, as the perturbation strengthens, the time it takes
for the particle to reach an equilibrium position decreases.  Also,
for the largest amplitudes studied, we note an almost immediate
offset from zero develops before the exponential behavior begins.
\label{unstabmult}}
\end{figure}

\begin{figure}
\scalebox{0.5}{
\includegraphics{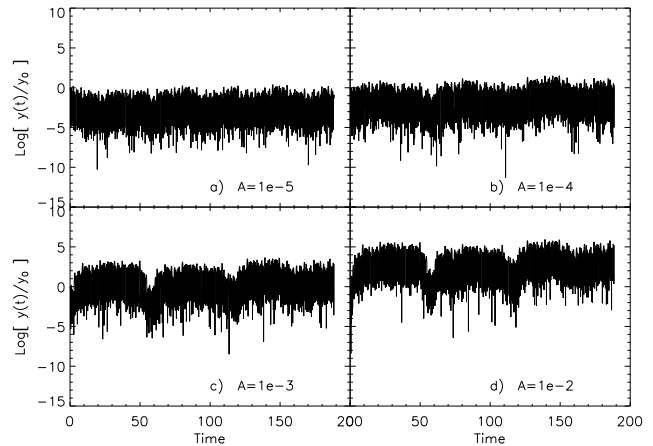}}
\caption{Behavior for an stable orbit in an gravitational potential
with triaxial values $a=5$, $b=4$, $c=3$.  The panels correspond to
the same increase in perturbation strength shown in
Figure~\ref{unstabmult}.  The signature of the Adams instability fails
to appear at any perturbing strength investigated.
\label{stabmult}}
\end{figure}

\begin{figure}
\scalebox{0.5}{
\includegraphics{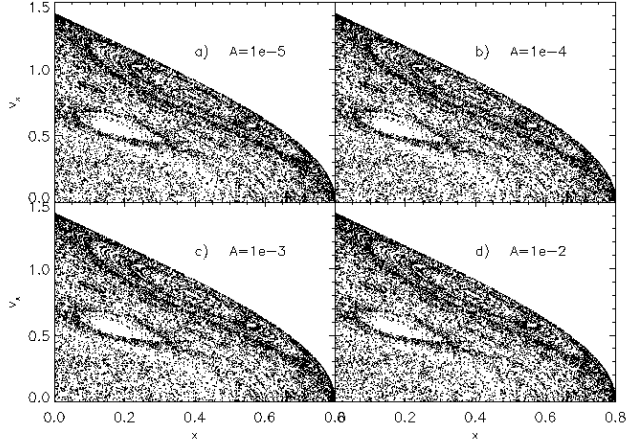}}
\caption{Panels a through d again correspond to the same increasing
perturbation strengths as in Figure~\ref{unstabmult}.  Each panel
shows a surface of section plot based on orbits in triaxial potentials
with $a=5$, $b=4$, $c=3$.  Orbits take on a range of initial $x$ and
$z$ positions, but $|y|_{\rm initial} <10^{-8}$.  We interpret the lack of
differences in the plots as evidence against the perturbers changing
the overall character of orbits.
\label{sosmult}}
\end{figure}

For the values of $A$ and $\sigma_p$ adopted, we find that orbits that
are stable in a smooth potential remain stable in a bumpy version.
Conversely, smooth-potential unstable orbits remain unstable.  We have
been unable to convert stable orbits to unstable and vice versa.  The
two specific orbits resulting in Figures~\ref{unstabmult} and
\ref{stabmult} illustrate this result.  A broader range of orbit
families have also been studied.  Figure~\ref{sosmult} shows various
surfaces of section plots based on these families.  The smooth
potential underlying these plots is the same as the one used to create
Figure~\ref{gridsos}.  The panels highlight the lack of substantial
changes as the perturbing strength changes.  We interpret these
results to mean that discreteness effects are not changing the basic
structure of phase space.  However, the time it takes an unstable
orbit to reach its steady-state $y$-position decreases as the strength
of the perturbations grows, as illustrated in Figure~\ref{unstabmult}.
The behaviors of the $y$-positions of orbits in our \textit{N}-body
simulations (\eg, Figure~\ref{nblogy}) are similar to those that
result from encounters with strong perturbers.  Specifically, the
extremely rapid rise followed by a slower exponential increase leading
to an eventual steady-state value is a common pattern.  This supports
the view that at least some of the motions away from the principal
plane that occur in \textit{N}-body simulations are caused by
discreteness and not the Adams instability. 

\begin{figure}
\scalebox{0.5}{
\includegraphics{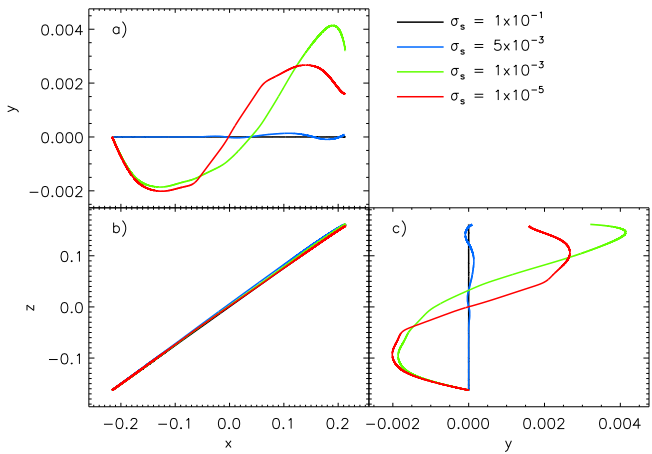}}
\caption{Projections of orbits ($x-y$, $x-z$, and $y-z$ in panels a,
b, and c, respectively) in four different potentials derived from a
spherical \textit{N}-body system.  Each orbit starts with the same
initial conditions.  The smoothing lengths used to create the various
potentials are given in the legend.  For comparison, the initial
system has a radius of one.  As expected, the impact of discreteness
decreases with increasing smoothing length.
\label{smoothorb}}
\end{figure}

In order to test the possibility that the overall potentials do not
have the appropriate shapes and/or profiles to support the Adams
instability, we have also investigated the behavior of orbits in
smoothed versions of our \textit{N}-body simulations.  Based on
particle positions corresponding to highly triaxial states, we use a
Gaussian kernel to smooth the mass distribution.  Each particle is
replaced by a Gaussian mass distribution with a prescribed scale length
$\sigma_s$.  This modifies that particle's contribution to the
potential at a point with displacement $\mathbf{r}$ by a term
proportional to the error function of $r/\sigma_s$.  Small $\sigma_s$
values (compared to the semiaxis lengths) result in
essentially point-like behavior.  For a given smoothed mass
distribution, a grid of potential and force values are calculated.
Again using the procedures in Section~\ref{adams}, a variety of
initial conditions have been numerically integrated.  As examples of
the impact of smoothing on the orbits, Figure~\ref{smoothorb} shows
segments of orbit paths derived from identical initial conditions in
potentials with different smoothing lengths.  The \textit{N}-body
model that serves as the basis for these potentials is spherical.  The
orbit is one that should run essentially through the center of the
potential along the $x$-axis, if it were perfectly smooth.  The
discrete nature of the relatively unsmoothed ($\sigma_s=10^{-5}$)
potential is evident.  

\begin{figure}
\scalebox{0.5}{
\includegraphics{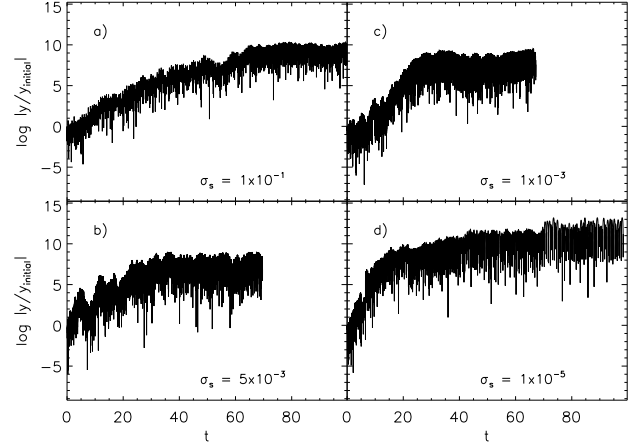}}
\caption{The $y$-positions of orbits in smoothed versions of a
triaxial \textit{N}-body potential.  Panels a, b, c, and d correspond
to smoothing lengths $\sigma_s=1\times 10^{-1}$, $\sigma_s=5\times
10^{-3}$, $\sigma_s=1\times 10^{-3}$, and $\sigma_s=1\times 10^{-5}$,
respectively.  The orbits all have the same initial conditions.  The
positions are normalized to the initial value, which is $\approx
10^{-8}$.  The exponential growth inherent with the Adams instability
is evident, and the orbits show the same qualitative behavior as those
in Section~\ref{adams}.  The major distinction between the panels is
that the steady-state $y$-position is reached more rapidly in less
smoothed potentials.
\label{nbasig}}
\end{figure}

Figure~\ref{nbasig} shows the signature of the Adams instability for a
particle orbiting in a highly smoothed ($\sigma_s=0.1$) version of a
triaxial $c/a\approx 0.5, b/a\approx 0.7$ \textit{N}-body potential.
In line with the previous findings of this section, we see the same
signature in potentials with smaller smoothing lengths, but the ``rise
time'' of the instability decreases with the smoothing length.

Overall, these findings suggest that numerical issues can conflate
physical processes to disguise the presence of the Adams instability.
The specific signature of the instability seen in smooth potentials
has its quantitative details altered by finite-\textit{N} effects, but
the qualitative behavior remains.

\section{Conclusions}\label{conclude}

Numerous previous studies have investigated how initially spherical
\textit{N}-body systems become triaxially distorted during
self-gravitating evolution.  The most common explanation involves what
has become known as the radial orbit instability.  Axes along which
more mass resides in these simulations can strongly influence the
tangential motions of neighboring orbits, causing alignments which
lead to non-sphericity.  However, these systems do not evolve to
extreme triaxiality, as one would expect if the radial orbit
instability operated continuously.  We have investigated a way in
which the growth of triaxiality in simulations of collapsing systems
could be halted.  Beginning with a previously examined orbital
instability that occurs in triaxial systems \citep{aetal07}, we have
shown that what we term the Adams instability persists beyond the
centrally-limited situation originally investigated.  The Adams
instability gives rise to a characteristic exponential growth in
position perpendicular to a principal axis.  In this way,
triaxial-supporting orbits can be depopulated.

To determine the presence of the Adams instability in simulations, we
have created and analyzed evolutions of \textit{N}-body systems from a
variety of initial conditions.  Following previous work, we control
the amount of initial random kinetic energy present in the systems to
create both spherical and triaxial evolutions.  Particle number,
softening length, and the presence of cosmological expansion have all
been varied for our investigation.  By tracking orbits near principal
planes of systems, we have seen the signature of the Adams
instability.  However, this signature also appears in systems that are
essentially spherical.  Individual orbits of particles making up
\textit{N}-body systems do not provide unambiguous evidence for the
Adams instability.

After creating perturbed versions of the smooth potentials used to
discover the Adams instability, we have found that small-scale noise,
like that due to discreteness effects in an \textit{N}-body
simulation, can make the onset of the instability more rapid.
However, the perturbations studied cannot change an orbit that would
be stable in a smooth potential into an unstable one, and vice versa.  
The behavior of orbits in sufficiently perturbed potentials does
mirror what is seen in our \textit{N}-body evolutions.

We have also looked for the Adams instability in smoothed versions of
our \textit{N}-body systems.  Fixed-time snapshots of our systems have
been used to create smoothed potentials.  Tracking orbital motions of
particular initial conditions as the smoothing length changes shows
that these potentials will support the Adams instability.  It appears
that a combination of discreteness effects and the Adams instability
are present in \textit{N}-body simulations like ours.  Numerical
effects can disguise the quantitative outcomes of the instability, but
they cannot erase its qualitative behavior.  Given that physical
systems should be essentially free of discreteness effects, the
inability to quantify the contributions of the two mechanisms to
simulated end-state shapes gives makes questionable any
simulation-based prediction of impact on the shapes of real-world
systems.

\acknowledgments
The authors thank the Wisconsin Space Grant Consortium for supporting
this work through the Undergraduate Research Fellowship program.


\begin{thebibliography}{99}

\bibitem[Adams \etal(2007)]{aetal07}
Adams, F.C., Bloch, A.M., Butler, S.C., Druce, J. M., Ketchum, J.A.
2007, \apj, 670, 1027
\bibitem[Allgood \etal(2006)]{aetal06}
Allgood, B., Flores, R.A., Primack, J.R., Kravtsov, A.V., Wechsler,
R.H., Faltenbacher, A., Bullock, J.S. 2006, \mnras, 367, 1781
\bibitem[Bailin \& Steinmetz(2005)]{bs05}
Bailin, J., Steinmetz, M. 2005, \apj, 627, 647
\bibitem[Bardeen \etal(1986)]{betal86}
Bardeen, J.M., Bond, J.R., Kaiser, N., Szalay, A.S. 1986, \apj, 304,
15
\bibitem[Barnes \& Efstathiou(1987)]{be87}
Barnes, J., Efstathiou, G. 1987, \apj, 319, 575
\bibitem[Barnes \etal(2005)]{betal05}
Barnes, E.I., Williams, L.L.R., Babul, A., Dalcanton, J.J. 2005, \apj,
634, 775
\bibitem[Barnes, Lanzel, \& Williams(2009)]{blw09} 
Barnes, E.I., Lanzel, P.A., Williams, L.L.R. 2009, \apj, 704, 372
\bibitem[Bellovary \etal(2008)]{betal08}
Bellovary, J.M., Dalcanton, J.J., Babul, A., Quinn, T.R., Maas, R.W.,
Austin, C.G., Williams, L.L.R., Barnes, E.I. 2008, \apj, 685, 739
\bibitem[Benhaiem \& Sylos Labini(2015)]{bs15}
Benhaiem, D., Sylos Labini, F. 2015, \mnras, 448, 2634
\bibitem[Binney(1981)]{b81}
Binney, J. 1981, \mnras, 196, 455
\bibitem[Binney \& Tremaine(1987)]{bt87}
Binney, J., Tremaine, S. 1987, {\it Galactic Dynamics} (Princeton, NJ:
Princeton Univ. Press)
\bibitem[Chandrasekhar(1969)]{c69}
Chandrasekhar, S. 1969, {\it Ellipsoidal Figures of Equilibrium} (New
Haven, CT: Yale Univ. Press)
\bibitem[Dubinski \& Carlberg(1991)]{dc91}
Dubinski, J., Carlberg, R.G. 1991, \apj, 378, 496
\bibitem[Frenk \etal(1988)]{fetal88}
Frenk, C.S., White, S.D.M., Davis, M., Efstathiou, G. 1988, \apj, 327,
507
\bibitem[Hernquist(1990)]{h90}
Hernquist, L. 1990, \apj, 356, 359
\bibitem[Jing \& Suto(2002)]{js02}
Jing, Y.P., Suto, Y. 2002, \apj, 574, 538
\bibitem[Merritt \& Aguilar(1985)]{ma85}
Merritt, D., Aguilar, L.A. 1985, \mnras, 217, 787
\bibitem[Merritt \& Fridman(1996)]{mf96}
Merritt, D., Fridman, T. 1996, \apj, 460, 136
\bibitem[Navarro, Frenk, \& White(1996)]{nfw96}
Navarro, J.F., Frenk, C.S., White, S.D.M. 1996, \apj, 462, 563
\bibitem[Palmer \& Papaloizou(1987)]{pp87}
Palmer, P.L., Papaloizou, J. 1987, \mnras, 224, 1043
\bibitem[Poon \& Merritt(2001)]{pm01}
Poon, M.Y., Merritt, D. 2001, \apj, 549, 192
\bibitem[Power \etal(2003)]{petal03}
Power, C., Navarro, J.F., Jenkins, A., Frenk, C.S., White, S.D.M., 
Springel, V., Stadel, J., Quinn, T. 2003, \mnras, 338, 14
\bibitem[Press \etal(1994)]{petal94}
Press, W.H., Teukolsky, S.A., Vetterling, W.T., Flannery, B.P. 1994,
{\it Numerical Recipes} (Cambridge, UK: Cambridge Univ. Press)
\bibitem[Shaya \etal(2010)]{setal10}
Shaya, E., Olling, R., Ricotti, M., Majewski, S.R., Patterson, R.J.,
Allen, R., van der Marel, R., Brown, W., Bullock, J., Burkert, A.,
Combes, F., Gnedin, O., Grillmair, C., Kulkarni, S., Guhathakurta, P.,
Helmi, A., Johnston, K., Kroupa, P., Lake, G., Moore, B., Tully, R.B.
2010, Astro2010: The Astronomy and Astrophysics Decadal Survey, 
Science White Papers, no. 274
\bibitem[Springel(2005)]{s05}
Springel, V. 2005, \mnras, 364, 1105

\end{thebibliography}
\end{document}